\documentclass[preprint,12pt,authoryear]{elsarticle}

\usepackage{amsmath,amsthm}
\usepackage{amssymb}
\usepackage{epstopdf}
\usepackage[caption=false]{subfig}
\usepackage{soul}
\usepackage{natbib}
\usepackage[colorlinks=true,urlcolor=blue,linkcolor=blue,citecolor=blue]{hyperref}
\usepackage{float}

\theoremstyle{definition}

\theoremstyle{remark}

\journal{arXiv}

\begin{document}

\begin{frontmatter}

\title{Eigenvalue-Based Randomness Test for Residual Diagnostics in Panel Data Models\tnoteref{articletype}}

\author[inst1]{Marcell T. Kurbucz\corref{cor1}}
\ead{m.kurbucz@ucl.ac.uk}
\cortext[cor1]{Corresponding author. ORCID: 0000-0002-0121-6781}

\author[inst2]{Betsab\'e P\'erez Garrido}
\author[inst3,inst4]{Antal Jakov\'ac}

\affiliation[inst1]{
  organization={Institute for Global Prosperity, The Bartlett, University College London},
  addressline={149 Tottenham Court Road},
  city={London},
  postcode={W1T 7NF},
  country={United Kingdom}
}
\affiliation[inst2]{
  organization={Department of Computer Science, Institute of Data Analytics and Information Systems, Corvinus University of Budapest},
  addressline={8 F\H{o}v\'am Square},
  city={Budapest},
  postcode={1093},
  country={Hungary}
}
\affiliation[inst3]{
  organization={Department of Computational Sciences, Institute for Particle and Nuclear Physics, HUN-REN Wigner Research Centre for Physics},
  addressline={29-33 Konkoly-Thege Mikl\'os Street},
  city={Budapest},
  postcode={1121},
  country={Hungary}
}
\affiliation[inst4]{
  organization={Department of Statistics, Institute of Data Analytics and Information Systems, Corvinus University of Budapest},
  addressline={8 F\H{o}v\'am Square},
  city={Budapest},
  postcode={1093},
  country={Hungary}
}

\begin{abstract}
This paper introduces the Eigenvalue-Based Randomness (EBR) test---a novel approach rooted in the Tracy-Widom law from random matrix theory---and applies it to the context of residual analysis in panel data models. Unlike traditional methods, which target specific issues like cross-sectional dependence or autocorrelation, the EBR test simultaneously examines multiple assumptions by analyzing the largest eigenvalue of a symmetrized residual matrix. Monte Carlo simulations demonstrate that the EBR test is particularly robust in detecting not only standard violations such as autocorrelation and linear cross-sectional dependence (CSD) but also more intricate non-linear and non-monotonic dependencies, making it a comprehensive and highly flexible tool for enhancing the reliability of panel data analyses.
\end{abstract}


\end{frontmatter}

\section{Introduction}
\label{S:1}

\noindent
Panel data models, which integrate both cross-sectional and time-series information, are fundamental in econometric analysis, enabling researchers to control for unobserved heterogeneity and better understand complex relationships within the data. These models rely on several key assumptions, including the exogeneity of the regressors, homoscedasticity (constant variance of residuals), and no autocorrelation (independence of residuals over time) \citep[see, e.g.,][]{meijer2017consistent}, which collectively ensure that residuals behave randomly without exhibiting any systematic patterns. When these assumptions are violated, parameter estimates can become biased or inefficient, leading to flawed inferences that undermine the reliability of the analysis \citep{garba2013investigations}.

To ensure robustness, various diagnostic tests are applied to assess the properties of residuals. Tests for cross-sectional dependence (CSD), such as the Lagrange Multiplier (LM) test \citep{breusch1980lagrange} and the CD test by \citeauthor{pesaran2004general} (\textcolor{blue}{2004}), evaluate whether residuals from different cross-sectional units are correlated. Non-linear monotonic relationships can also be detected using methods such as Frees' test \citep{frees1995assessing} and recent rank-correlation techniques by \cite{feng2021rank}. Autocorrelation is examined with tests such as the Breusch-Godfrey test \citep{breusch1978testing, godfrey1978testing} and the LM test by \citeauthor{baltagi1991joint} (\textcolor{blue}{1991}), which address inefficiencies from residuals that are not independent over time. Heteroscedasticity is assessed using methods like the Breusch-Pagan test \citep{breusch1979simple}, the White test \citep{white1980heteroskedasticity}, and the Modified Wald test \citep{greene2008econometric} for groupwise heteroscedasticity in panel data. Normality is evaluated through tests such as those by \cite{jarque1980efficient} and \cite{shapiro1965analysis}, which help identify potential specification errors, while the Ramsey RESET test \citep{ramsey1969tests} detects functional form misspecifications. Finally, ensuring stationarity with unit root and stationarity tests \citep[see, e.g.,][]{levin2002unit, im2003testing, hadri2008panel} is crucial to avoid spurious results in panel data analysis.

This paper introduces a novel diagnostic approach, the Eigenvalue-Based Randomness (EBR) test, which, unlike the aforementioned tests, simultaneously assesses multiple assumptions while directly testing the randomness of the residual matrix. The approach utilizes the Tracy-Widom law \citep{tracy1996orthogonal}---a concept widely applied in physics \citep[see, e.g.,][]{bajnok2024tracy, betea2024multicritical} but largely unexplored in econometrics---which characterizes the asymptotic behavior of the largest eigenvalue of random matrices. By comparing the largest eigenvalue of the symmetrized residual matrix against the Tracy-Widom distribution \citep{tracy1996orthogonal, tracy2009distributions}, this new test offers a comprehensive and robust evaluation of residual behavior, capable of detecting systematic patterns---including non-monotonic CSD---that traditional methods may overlook.

The rest of this paper is organized as follows: Section~\ref{S:2} details the proposed testing procedure; Section~\ref{S:3} describes the dataset used in the study; Section~\ref{S:4} presents and discusses the empirical findings; and finally, Section~\ref{S:5} offers conclusions and outlines future research directions.

\section{Eigenvalue-Based Randomness Test}
\label{S:2}

\noindent
In this study, we assess the randomness of the error term in panel econometric models by analyzing the largest eigenvalue of the residual matrix. The methodology involves the following steps:

\vskip 0.3cm
\noindent
1) \underline{Identifying the Largest Eigenvalue:}
\vskip 0.3cm

\noindent
First, let us consider $\mathbf{E}$, an $m \times n$ dimensional residual matrix obtained from the panel econometric model and its standardized form:

\begin{equation}
\mathbf{E}^* = \frac{\mathbf{E} - \mu}{\sigma},
\end{equation}

\noindent
where $\mu$ is the mean of all elements in $\mathbf{E}$, and $\sigma$ is their standard deviation. To facilitate the application of random matrix theory, as a next step, we construct a square matrix $\mathbf{Z}$ by supplementing elements of $\mathbf{E}^*$ with elements drawn from a standard normal distribution \(\mathcal{N}(0, 1)\). Formally, the square matrix \(\mathbf{Z}\) is constructed as:

\begin{equation}
\mathbf{Z} = 
\begin{cases}
\bigl[\mathbf{E}^* \mid \mathbf{G}_{m \times (m - n)}\bigr], & \text{if } m > n, \\[6pt]
\begin{bmatrix}
\mathbf{E}^* \\
\mathbf{H}_{(n - m) \times n}
\end{bmatrix}, & \text{if } m < n, \\[6pt]
\mathbf{E}^*, & \text{if } m = n,
\end{cases}
\end{equation}

\noindent
where $\mathbf{G}_{m \times (m - n)}$ is an $m \times (m - n)$ matrix, and $\mathbf{H}_{(n - m) \times n}$ is an $(n - m) \times n$ matrix, with each element independently drawn from the standard normal distribution $\mathcal{N}(0, 1)$. To obtain real eigenvalues, we symmetrize $\mathbf{Z}$ as follows:

\begin{equation}
\mathbf{S} = \frac{1}{2} \left( \mathbf{Z} + \mathbf{Z}^\top \right),
\end{equation}

\noindent
and then perform the spectral decomposition of the symmetric matrix $\mathbf{S}$. This step yields eigenvalues $\lambda_1, \lambda_2, \dots, \lambda_k$, where $\lambda_1 \geq \lambda_2 \geq \dots \geq \lambda_k$ and $k = \max(m, n)$. Our further analysis focuses on the largest eigenvalue $\lambda_1$ of the matrix $\mathbf{S}$.

\vskip 0.3cm
\noindent
2) \underline{Testing the Largest Eigenvalue:}
\vskip 0.3cm

\noindent
After the spectral decomposition, we conduct a statistical test to determine whether the largest eigenvalue $\lambda_1$ follows the Tracy-Widom distribution, denoted by $\text{TW}_\beta$. The Tracy-Widom distribution is appropriate for the largest eigenvalue in certain types of random matrices under large-dimensional asymptotics. The null hypothesis is:

\begin{equation}
H_0: \lambda_1 \sim \text{TW}_\beta,
\end{equation}

\noindent
where the order $\beta = 1$ is based on the assumption that the residual matrix is real and symmetric, corresponding to the Gaussian orthogonal ensemble (GOE). By testing this hypothesis, we assess the randomness of the error term in the panel econometric model. A significant deviation from the Tracy-Widom distribution would suggest that the residuals exhibit non-random structures, indicating potential model misspecification or other underlying issues.

\section{Monte Carlo Simulation} \label{S:3}

\begin{sloppypar}
\noindent
To investigate the properties of the proposed residual diagnostic method, a series of Monte Carlo experiments were conducted. The simulation procedure involved generating matrices with dimensions of $n \times m$, where $n\in\small\{30,50,100\small\}$ represents the number of cross-sectional units and $m\in\small\{15,20,50\small\}$ denotes the number of time periods.\footnote{For reference, we utilized similar combinations of cross-sectional units and time periods as those in \cite{feng2021rank}.} For each parameter setup, $1{,}000$ matrices were simulated, and the power of the proposed EBR test---defined as the probability of correctly rejecting the null hypothesis when it is false---was assessed at a significance level of $\alpha = 5\%$.
\end{sloppypar}

The simulations aimed to explore the effects of autocorrelation, linear CSD, and non-monotonic dependence among the residuals across different cross-sectional units. These three experiments were conducted as follows.

\vskip 0.3cm
\noindent
a) \underline{Autocorrelation Case:}
\vskip 0.3cm

\noindent
In the autocorrelation case simulations, matrices are initially generated as random with no CSD. Temporal structure is introduced by applying an AR(1) process to each time series:

\begin{equation}
x_{i,t} = \phi \cdot x_{i,t-1} + (1 - \phi) \cdot \epsilon_{i,t},
\end{equation}

\noindent
where $\phi$ is the autocorrelation coefficient, and $\epsilon_{i,t}$ represents the initial residuals. This process simulates temporal persistence within each cross-sectional unit.

\vskip 0.3cm
\noindent
b) \underline{Linear CSD Case:}
\vskip 0.3cm

\noindent
In the linear CSD case, matrices were generated with a predefined correlation between cross-sectional units. A covariance matrix $\Sigma$ was constructed with elements $\sigma_{ij}$ where:

\begin{equation}
\sigma_{i,j} = 
\begin{cases} 
\rho & \text{for } i \neq j, \\
1 & \text{for } i = j,
\end{cases}
\end{equation}

\noindent
where $\rho$ represents the strength of the linear CSD. Residuals were drawn from a multivariate normal distribution $\mathcal{MVN}(\mathbf{0}, \Sigma)$, ensuring linear dependence across units.

\vskip 0.3cm
\noindent
c) \underline{Non-Monotonic CSD Case:}
\vskip 0.3cm

\noindent
The non-monotonic CSD case involved generating independent random data. Complex non-linear transformations were then applied to half of the columns, as follows:

\begin{equation}
x_{i,t} = \sin(x_{i,t-1}) + \cos(x_{i,t-1}^2) + 0.5 \cdot \epsilon_{i},
\end{equation}

\noindent
where $\epsilon_{i}$ represents random noise,  drawn from a standard normal distribution $\mathcal{N}(0, 1)$. This created non-monotonic relationships among the cross-sectional units, capturing more intricate dependencies than the linear case.

\section{Results and Discussion}
\label{S:4}

\begin{figure}[H]
\caption{Power of the EBR test in the presence of autocorrelation}
\vspace{0.3em}
\label{fig:1}
\centering
\includegraphics[width=0.715\textwidth]{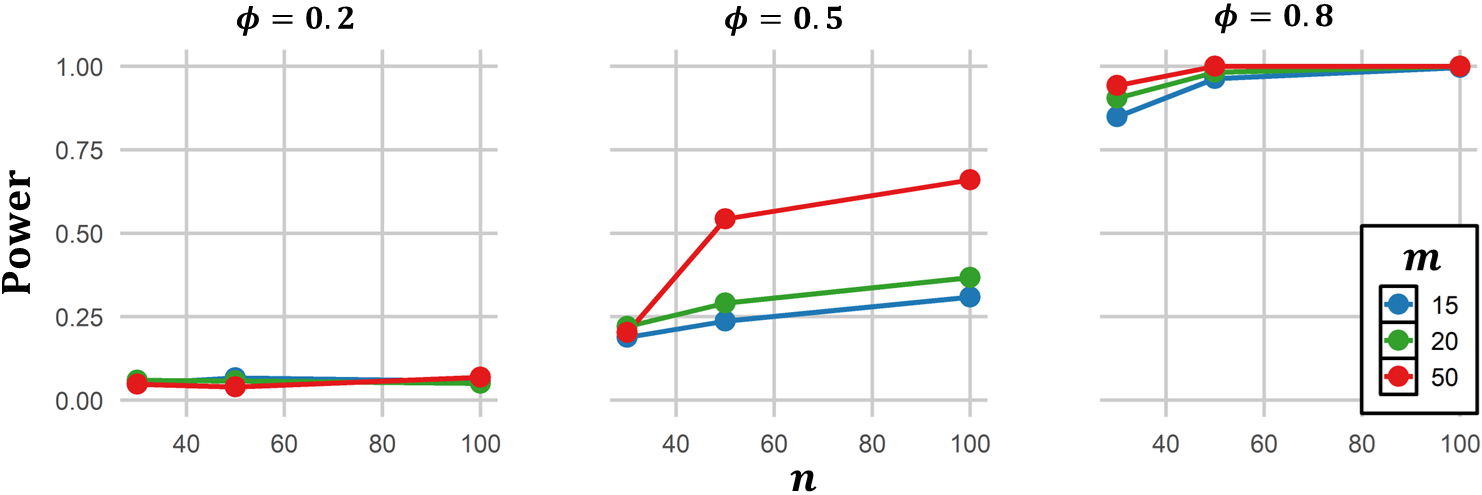}
\end{figure}

\begin{figure}[H]
\caption{Power of the EBR test in the presence of linear CSD}
\vspace{0.3em}
\label{fig:2}
\centering
\includegraphics[width=0.715\textwidth]{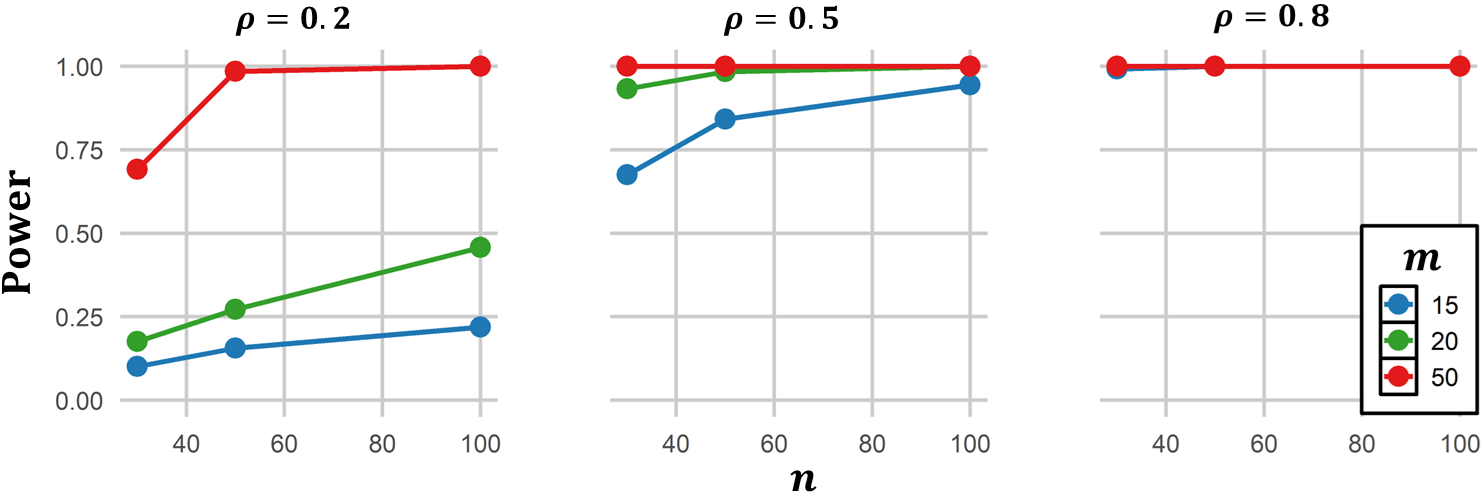}
\end{figure}

\begin{figure}[H]
\caption{Power of the EBR test in the presence of non-monotonic CSD}
\vspace{0.3em}
\label{fig:3}
\centering
\includegraphics[width=0.715\textwidth]{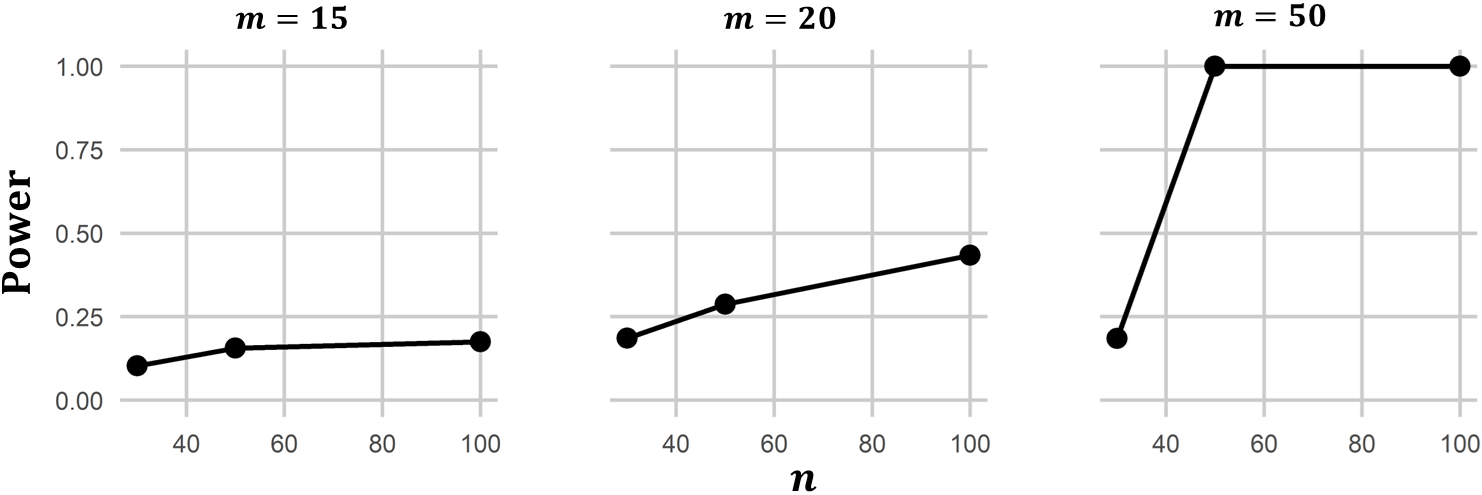}
\end{figure}

\noindent
The outcomes of the Monte Carlo experiments are presented in Figures \ref{fig:1} through \ref{fig:3}, followed by a subsequent discussion.

\vskip 0.3cm
\noindent
a) \underline{Autocorrelation Case:}

\noindent
As depicted in Figure 1, the EBR test was unable to detect weak autocorrelation ($\phi = 0.2$) under the current settings. However, in the case of high autocorrelation ($\phi = 0.8$), the test demonstrates a power close to $1$, even with short time periods. For moderate autocorrelation ($\phi = 0.5$), the detection capability of the test is significantly influenced by the number of time periods included in the panel model.

\vskip 0.3cm
\noindent
b) \underline{Linear CSD Case:}

\noindent
The proposed method effectively detects linear CSD, even in cases of weak CSD ($\rho = 0.2$) and relatively small panel models. In such settings, the test approaches its theoretical maximum power with as few as $50$ cross-sectional units and $50$ time periods. For stronger CSD effects, the method achieves robust power regardless of the model size.

\vskip 0.3cm
\noindent
c) \underline{Non-Monotonic CSD Case:}

\noindent
While traditional, Pearson correlation-based CSD tests are effective in detecting linear dependencies, they fall short in identifying non-monotonic relationships between cross-sectional units. Even alternative methods based on rank correlation fail in these cases.\footnote{In this experiment, the lower triangle of the Pearson correlation matrix showed a mean of $0.018$ and a standard deviation of $0.027$. For Kendall and Spearman correlations, the means were $0.002$ and $0.003$, with standard deviations of $0.010$ and $0.015$, respectively.} In contrast, the proposed EBR test is well-suited for such scenarios, achieving its maximal theoretical power with as few as $50$ cross-sectional units and $50$ time periods. This result highlights the effectiveness of the randomness tests in detecting complex dependencies among cross-sectional units.

\section{Conclusions and Future Works}
\label{S:5}

\noindent
The EBR test introduced in this paper offers a significant enhancement to the suite of diagnostic tools available for panel data analysis. By leveraging the Tracy-Widom distribution to assess the largest eigenvalue of the symmetrized residual matrix, this test provides a comprehensive evaluation of residual behavior, effectively identifying complex and subtle patterns that traditional diagnostic methods may overlook. The Monte Carlo simulations validate the test's robustness across a range of scenarios, including high levels of autocorrelation, linear cross-sectional dependence, and non-monotonic relationships among residuals. The EBR test's ability to simultaneously address multiple assumptions about residual randomness marks a substantial improvement in ensuring the reliability of econometric inferences.

Future research could involve an in-depth comparison of the EBR test with traditional diagnostic methods across various scenarios, offering a clearer understanding of its advantages and limitations. Additionally, investigating the joint effects of multiple diagnostic cases, such as how autocorrelation and cross-sectional dependence interact within panel data, could further refine this approach's applicability. In the financial sector, the EBR test could be applied to model systemic risk among interconnected financial institutions, where it may uncover complex, non-linear dependencies missed by traditional diagnostics, leading to better risk assessments and more informed regulatory decisions. Moreover, the EBR test offers significant promise in spatial econometrics, particularly for analyzing spatial cross-sectional dependence (CSD) tied to location, and may also prove especially valuable for gravity model testing. In the latter context, validation often involves generating synthetic networks from estimated models and comparing their properties with observed data. By contrast, the EBR test could capture complex dependencies in the residual matrix of a gravity model, potentially offering a more precise and efficient approach to assessing model validity.

\section*{Acknowledgments}

\noindent
Supported by the ÚNKP-23-4-II-CORVINUS-11 New National Excellence Program of the Ministry for Culture and Innovation from the source of the National Research, Development and Innovation Fund.

\section*{Data Availability}

\noindent
The data supporting the findings of this study are available from the corresponding author upon reasonable request.

\section*{Disclosure Statement}

\noindent
The authors report there are no competing interests to declare.

\bibliographystyle{apalike}
\bibliography{cas-refs}

\end{document}